\documentclass[twocolumn]{aastex63}
\usepackage{amssymb}	
\usepackage{amsmath}	
\usepackage{lineno}
\received{June 1, 2019}
\revised{January 10, 2019}
\accepted{\today}
\submitjournal{ApJ}

\shorttitle{CSM around SNe~Ia}
\shortauthors{M.-K. Hu et al.}
\graphicspath{{./}{figures/}}

\begin{document}

\title{The Effects of Circumstellar Dust Scattering on the Light Curves and Polarizations of Type Ia Supernovae\footnote{Supported by the National Natural Science Foundation of China.}}

\email{lifan@tamu.edu} 
\email{kaihukaihu123@pmo.ac.cn} 

\author{Maokai Hu}
\affiliation{ Purple Mountain Observatory, Chinese Academy of Sciences, Nanjing 210023, China}
\affiliation{ School of Astronomy and Space Science, University of Science and Technology of China, Hefei, Anhui 230026, China}

\author{Lifan Wang}
\affiliation{ George P. and Cynthia Woods Mitchell Institute for Fundamental Physics \& Astronomy, Texas A\&M University, Department of Physics and Astronomy, 4242 TAMU, College Station, TX 77843, USA}

\author{Xiaofeng Wang} 
\affiliation{Physics Department and Tsinghua Center for Astrophysics (THCA), Tsinghua University, Beijing, 100084, China}
\affiliation{Beijing Planetarium, Beijing Academy of Science and Technology, Beijing, 100044, China}



\begin{abstract}

Observational signatures of the circumstellar material (CSM) around Type Ia supernovae (SNe~Ia) provide a unique perspective on their progenitor systems. The pre-supernova evolution of the SN progenitors may naturally eject CSM in most of the popular scenarios of SN~Ia explosions. In this study, we investigate the influence of dust scattering on the light curves and polarizations of SNe~Ia. A Monte Carlo method is constructed to numerically solve the process of radiative transfer through the CSM. Three types of geometric distributions of the CSM are considered: spherical shell, axisymmetric disk, and axisymmetric shell. We show that both the distance of the dust from the SN and the geometric distribution of the dust affect the light curve and color evolutions of SN. We found that the geometric location of the hypothetical circumstellar dust may not be reliably constrained based on photometric data alone, even for the best observed cases such as SN~2006X and SN~2014J, due to the degeneracy of CSM parameters. Our model results show that a time sequence of broadband polarimetry with appropriate time coverage from a month to about one year after explosion can provide unambiguous limits on the presence of circumstellar dust around SNe~Ia. 

\end{abstract}



\section{Introduction} 

Type Ia supernovae (SNe~Ia) have well-defined light curves and are employed empirically as cosmological distance indicators \citep{1998AJ....116.1009R,2007ApJ...659...98R,1999ApJ...517..565P,WangL:2003ApJ...590..944W,He2018ApJ...857..110H}. Of particular interest is the nature of their progenitor systems (e.g., \citealt{2011NatCo...2..350H,2014ARA&A..52..107M}). Theoretically there are two major channels, and both involve white dwarfs (WDs) in binary systems (e.g., \citealt{2000ARA&A..38..191H}). In the single degenerate channel the WD accretes matter from a nondegenerate star to reach the critical mass for SN explosion \citep{1973ApJ...186.1007W,1982ApJ...253..798N}, whereas in the double degenerate channel the explosion is achieved by the merging of the WD with a degenerate companion \citep{1984ApJS...54..335I,1984ApJ...277..355W}. In either case, circumstellar material (CSM) may be ejected before the explosion, and studies of this may provide unique clues to the nature of the progenitors of SNe~Ia \citep{2012ApJ...754L..21F,2013ApJ...770L..35S,2017ApJ...834...60Y,2019ApJ...882...30L,Ding2021}.

SN~2002ic is the first SN Ia found to show a strong ejecta$-$CSM interaction \citep{Hamuy2003Natur.424..651H,Wang2004ApJ...604L..53W,Wood-Vasey:2004ApJ...616..339W}. The SN~2002ic-like SNe Ia are identified by a spectroscopic transition from Type Ia to Type IIn after explosion. More such objects have been found  \citep{Aldering:2006ApJ...650..510A,2007ApJ...659L..13O,2012A&A...545L...7T,2015MNRAS.447..772F,2016MNRAS.459.2721I}. Further evidence of the presence of a significant amount CSM around SNe Ia came from spectroscopic observations of the narrow Na I D lines. Some SNe Ia show blueshifted and time-evolving narrow Na I D absorption lines \citep{2007Sci...317..924P,2009ApJ...693..207B,2009ApJ...702.1157S,2011Sci...333..856S,2013MNRAS.436..222M,Wang:2019ApJ...882..120W}. In particular, \citet{2009ApJ...699L.139W} divided the spectroscopic normal SNe~Ia into two groups: the normal-velocity ones and high-velocity ones, with Si II $\lambda$6355 velocity lower or higher than 11,800 $\text{km}\ \text{s}^{-1}$ respectively.  \cite{Wang:2019ApJ...882..120W} found that the SNe~Ia with high-speed Si II features tend to be systematically associated with blueshifted Na I D lines. According to these studies, the distances of the CSM from the SNe range from $10^{15}$ cm to $10^{19}$ cm, and the mass loss rates that lead to such CSM are usually lower than $10^{-4}-10^{-9}\ M_{\odot}\ \text{yr}^{-1}$ if they are the results of steady stellar winds, consistent with the constraints set by X-ray and radio observations \citep{2014ApJ...790...52M,2014ApJ...792...38P,2016ApJ...821..119C,2020ApJ...890..159L}. 

The presence of CSM can also alter the light curves and polarization of SNe~Ia, due to light echoes caused by dust scattering \citep{1986ApJ...308..225C,1996ApJ...462L..27W,2005MNRAS.357.1161P,2005ApJ...635L..33W,2008ApJ...686L.103G,Ding2021}. Light echoes from interstellar dust have been observed, such as the light echoes of SN~2006X \citep{2008ApJ...689.1186C,2008ApJ...677.1060W}, SN~2014J \citep{2015ApJ...804L..37C,2017ApJ...834...60Y}, and some supernova remnants \citep{2008ApJ...680.1137R,2012Natur.482..375R}. \cite{Bulla2018MNRAS.479.3663B} adopted a thin shell structure to fit the color evolution of several SNe~Ia in the context of dust scattering, and suggested that the shells are typically located at several parsecs away from the SNe. The result, however, as we will show in this study, is dependent on the assumed geometry of the dust distribution. \cite{2018MNRAS.476.4806N} studied the polarization of SN~2012hn with two asymmetric CSM geometries (disk-like and jet-like), where the degree of polarization may be as large as a few percent. Although the high degree of polarization predicted in \cite{2018MNRAS.476.4806N} is inconsistent with observations to date, such as those of SN~2005ke \citep{Patat:2012A&A...545A...7P}, 2009dc \citep{Tanaka:2010ApJ...714.1209T}, and 2014J \citep{Kawabata2014ApJ...795L...4K,2016ApJ...828...24P,Yang2018ApJ...854...55Y}, it does provide a way of identifying the geometric distribution of CSM. \cite{Yang2018ApJ...854...55Y} obtained precise polarization images of SN~2014J from $\sim277$ days to $\sim1181$ days after the maximum light, and the polarization signal can be modeled  by a dusty blob located at around $5\times10^{17}$ cm from the SN in the plane of the sky at the location of the SN. 

Monte Carlo (MC) simulations can be used to solve the dust scattering process (e.g., \citealt{Witt1977ApJS...35....1W,2001ApJ...551..269G,2013ARA&A..51...63S,Ding2021}). One application of this method is to simulate the polarization in dusty galaxies by virtue of the dust scattering through the interstellar material \citep{1996ApJ...465..127B,2013A&A...550A..74D,2017A&A...601A..92P}. Another example is the scattering by the CSM around core-collapse supernovae, where light echoes and polarization signals are calculated by the MC method \citep{2017ApJ...834..118M,2017ApJ...847..111N,Ding2021}. The Henyey$-$Greenstein phase function is usually used as the formula for dust scattering \citep{1941ApJ....93...70H}. Other dust properties, such as the albedo, the cross section, and the asymmetry factor, can be taken from \cite{Draine1984ApJ...285...89D} and \cite{Draine:2003ApJ...598.1017D}, assuming the dust properties are similar to those in either the Milky Way dust or the dust in the Large Magellanic Cloud. 

A set of models are presented in this paper for the scattering by the circumstellar dust of different geometric shapes around SNe~Ia. Because there is strong evidence that the dust around SNe~Ia may be systematically different from that in the Milky Way or the Large Magellanic Cloud \citep{WangL:2003ApJ...590..944W,Patat:2012A&A...545A...7P,Wang:2019ApJ...882..120W}, the dust properties are numerically calculated through Mie scattering theory for a given grain size distribution using the refractive index of \citet{Draine:2003ApJ...598.1017D}.  Section~\ref{model} describes the model, including the dust properties, the MC models, and the geometric distributions of the CSM. In Section~\ref{sec_result}, models are shown for a set of CSM distributions. Section~\ref{discussion} provides further discussions of the models and their applications to observational data. The conclusions are given in Section~\ref{conclusion}. 


\section{Models}  
\label{model}  

\subsection{Overview of the Radiative Transfer Process}

Generally, the process of radiative transfer through the circumstellar (CS) dust includes scattering, absorption, and re-emission. The re-emission contributes to infrared flux and will not be considered here. The photon state in the Monte Carlo process is described by the Stokes parameters ($S = (I, Q, U, V)^{\text{T}}$) following \citet{1950ratr.book.....C}, where $I$ is the intensity, $Q$ and $U$ describe linear polarization, $V$ describes circular polarization, and T stands for matrix transpose. The degree of linear polarization ($P$) can be written as $P = \sqrt{(Q^2 + U^2)} / I$, in which the circular polarization ($V$) is ignored in our models. Solving the radiative transfer process can be regarded as determining a kernel function that links the Stokes parameters before and after the photon$-$CSM interaction:
\begin{equation}
\label{eq11}
\begin{split}
S_\lambda(t', \Omega, \bar{U}) & = S_{\lambda,0}(t')\exp(-\tau_\lambda) \\
    & + \int I_{\lambda,0}(t'-t)K_{\lambda,S}(t, \Omega, \bar{U}) dt 
\end{split}
\end{equation}
where $t'$ is the time after explosion, $\Omega$ is the solid angle to the observer, $S_{\lambda,0}(t')$ is the Stokes parameter at wavelength $\lambda$ of the SNe~Ia before dust scattering, $\tau_\lambda$ is the optical depth at wavelength $\lambda$, $K_{\lambda,S}(t, \Omega, \bar{U})$ is a kernel function that can be calculated by assuming a $\delta$-function pulse as the input signal with $\bar{U}$ being an array describing the parameters related to the geometric distribution and optical properties of the dust. Equation~\ref{eq11} contains two parts: the transmitted component along the line of sight $S_{\lambda,0}(t')\exp(-\tau_\lambda)$, and the scattered component $\int I_{\lambda,0}(t'-t)K_{\lambda,S}(t, \Omega, \bar{U}) dt$. We will use the optical depth in the $B$ band as a measure of the optical properties of the CS dust. The optical depth of any given band can be directly calculated from that of the $B$ band based on Mie scattering for a given dust distribution. The kernel function $K_{\lambda,S}(t, \Omega, \bar{U})$ is a function of the dust properties, the scattering process, and the geometric distribution of CSM. 

\subsection{Dust Properties}
\label{sec_dust_mie}

In this study, all the values of albedo ($\omega$), scattering cross section ($\sigma_{\text{sca}}$), extinction cross section ($\sigma_{\text{ext}}$), and scattering matrix are numerically calculated from Mie scattering theory \citep{2004CoPhC.162..113W} based on the refractive index of dust grains from \cite{Draine:2003ApJ...598.1017D}. The size distribution of the dust grains takes the following form: 
\begin{equation}
f(r) = r^{-a_0}\exp\{-b_0(\log\frac{r}{r_0})^{2.0}\} 
\label{eq00}
\end{equation}
where $a_0$ and $b_0$ are $4.0$ and $7.5$, respectively. The shape of the curve given in Equation~\ref{eq00} is consistent with the results in \citet{2015ApJ...811L..39N}, with $r_0 = 0.05\ \mu m$ representing the small size of dust grains with average radius of $0.045\ \mu m$. The dust grains on the line of sight to SNe~Ia are likely to be smaller than typical dust grains in the Milky Way, as may be inferred from the low values of the ratio of total to selective extinction for typical SNe~Ia \citep{2005ApJ...635L..33W,2006ApJ...645..488W,2014MNRAS.443.2887F,2015MNRAS.453.3300A,2020P&SS..18304627G}. In addition, only  silicate grains with single chemical composition are considered in our models; the difference is insignificant for models with both silicate and graphite grains \citep{2015ApJ...807L..26G}.

\subsection{Monte Carlo Method} 
\label{subsectionMCcode}

The MC method includes several steps: the launching of photons, the tracking of photons through the CSM, and the integration of photons that have escaped from the CSM to build the kernel functions (Equation~\ref{eq11}) and solutions. Photons are launched with a given Stokes parameters in a specific direction and propagate a certain distance until being absorbed or scattered. The photons are assumed to be unpolarized initially, as can be justified by spectropolarimetry of SNe~Ia \citep{doi:10.1146/annurev.astro.46.060407.145139}, and their Stokes parameter is expressed as $(1,0,0,0)^{\text{T}}$. The geometric size of the SN is much smaller than the extent of the scattering material and is thus set to zero in all the calculations. The radiation from the SN is assumed to be spherically symmetric. The distance to the first photon$-$matter interaction depends on the optical depth in the radial direction, which is related to the composition and number density ($N(R)$) of dust grains. Assuming a steady stellar wind with constant velocity, the density can be described by $N(R) = A/R^2$ with $A$ being a scaling parameter and $R$ being the distance from the SN. The probability of a photon propagating a distance less than $R_{\text{in}} + D$ without interacting with a dust particle is expressed as $p(R<R_{\text{in}}+D) = 1 - \exp(-\tau(R_{\text{in}}+D)) $, where $R_{\text{in}}$ is the inner boundary of the CSM and $\tau(R_{\text{in}}+D)$ is the optical depth at the distance of $R_{\text{in}} + D$. The probability $p$ has a uniform distribution ranging from $0$ to $1-\exp(-\tau_0)$, where $\tau_{0}$ is the optical depth of CSM in the direction of photon propagation. This treatment of the scattering process is identical to that of \cite{Witt1977ApJS...35....1W}. Hence, the first free propagation distance $D$ in the CSM could be generated through an MC process: 
\begin{equation} 
D = \frac{R_{\text{in}}A\sigma_{\text{ext}}}{A\sigma_{\text{ext}} + R_{\text{in}}\ln(\epsilon)} - R_{\text{in}} 
\end{equation}
where $\epsilon$ is a random number in the range $(\exp(-\tau_0), 1)$. For scattering after the first interaction, we adopted  the same approach as \cite{Witt1977ApJS...35....1W} by assuming a locally uniform distribution of CSM; the propagation distance of a photon is expressed as $D = -\ln(\epsilon) / (N(R) \sigma_{\text{ext}}$), with the range of the random number $\epsilon$ being from $0$ to $1.0$.

The scattering process is calculated by computing the scattering angle following a distribution related to the scattering matrix and the Stokes parameters of the scattered photon by the rotational matrix and scattering matrix. To increase the computational efficiency, the absorption process is modeled by the weighting function as described in \cite{Witt1977ApJS...35....1W}. Once the photon is out of the CSM, the Stokes parameters are integrated to the same arrival time at the observer inside a solid angle interval $\Delta\Omega$. 

With the total number of photons ($N_{\text{photon}}$) emitted in the MC program, the kernel function of the Stokes parameter is reconstructed as
\begin{equation}
K_{\lambda,S}(t, \Omega, \bar{U}) = N_{\lambda,S}(t, \Omega,\bar{U} ) / N_{\text{photon}} \cdot (\Omega_{\text{emit}} / \Delta\Omega) 
\end{equation}
where $N_{\lambda,S}(t, \Omega, \bar{U})$ is the corresponding values of the Stokes parameter with a time delay of $t$ and integrated over the solid angle $\Delta\Omega$. The size of $\Delta\Omega$ determines the angular resolution of the model, $N_{\text{photon}}$ is the total number of injected photons in the calculations, $\Omega_{\text{emit}}$ is the solid angle in which the photons are injected into the CSM, and $\Delta\Omega$ is the solid angle over which the photons escaping from the CSM are integrated. For a spherically symmetric structure, the solid angles of both emitted and collected photons ($\Omega_{\text{emit}}$ and $\Delta\Omega$) are $4\pi$. For an axially symmetric disk or axisymmetric shell $\Delta\Omega$ is  $2\pi\sin\theta\Delta\theta$, where $\Delta\theta$ is the opening angle from the line of sight and is equal to $1^{\circ}$ in our model to ensure the accuracy of light curves and polarization. For convenience, the kernel function $K_{\lambda,S}(t, \Omega, \bar{U})$ is simplified to $K_S(t)$. $K_I(t)$, $K_Q(t)$, and $K_U(t)$ represent the kernel functions of Stokes parameters $I$, $Q$, and $U$, respectively. With all the reconstruction above, it is clear that $K_I = 1$ if there is no CSM-induced polarization.

\begin{figure}
\centering
\includegraphics[width = 0.95 \linewidth]{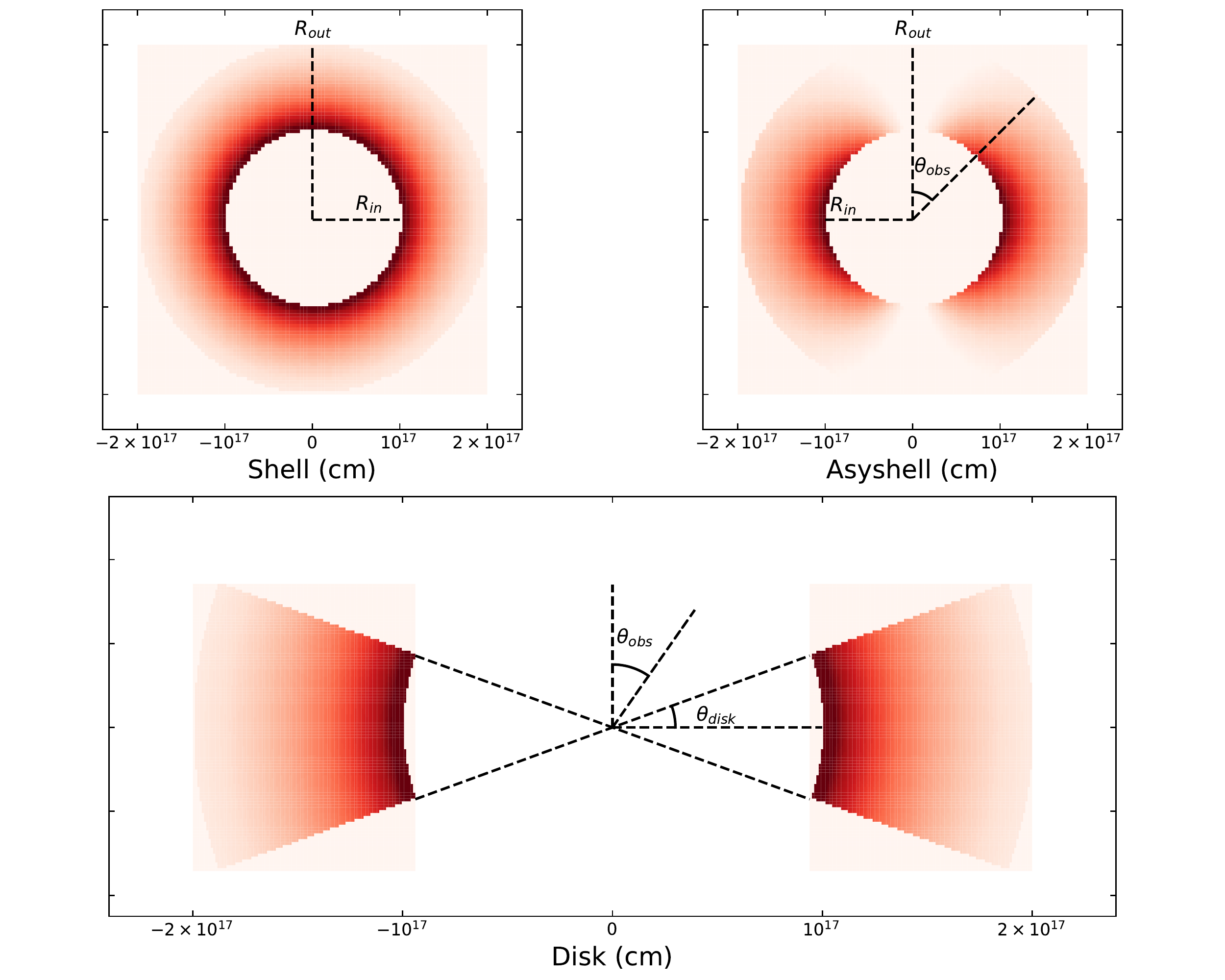}
\caption{Three types of CSM distributions: spherical shell (top left), axisymmetric shell (Asyshell, top right), and axisymmetric disk (bottom). The transparency of the color represents the number density of dust grains.} 
\label{fig_ilus_csm}
\end{figure}

\subsection{Geometric Distributions of CSM} 

We considered three different  geometric distributions of the CSM. These are spherical shells, axisymmetric disks, and axisymmetric shells --- as shown in Figure~\ref{fig_ilus_csm}, which is similar to the plot in Figure $3$ of \citet{Wang:2019ApJ...882..120W}. The shell or disk structures may arise from the stellar wind or accretion/excretion disks of the progenitor systems of SNe~Ia. The details of the geometric structure are not known but it is nonspherical, which can be expected based on observations of the stellar environment around known white dwarfs; the geometry carries important information in understanding the mass loss history of the progenitor systems. 

With Figure~\ref{fig_ilus_csm} we can define the model parameters for the calculations of dust scattering. These are the inner ($R_{\text{in}}$) and outer ($R_{\text{out}}$) radii that define the boundaries of the dust distribution; from them we define the extent of the CSM as $R_{\text{wid}} = R_{\text{out}} - R_{\text{in}}$. As shown in Figure~\ref{fig_ilus_csm}, the angle to the observer is given by $\theta_{\text{obs}}$, and the opening angle of the disk is $\theta_{\text{disk}}$. For the shell and disk structures, the number density of dust grains in the radial direction is given as $A/R^2$, where the parameter $A$ is a scaling constant and $R$ is the distance to the SN. For the axisymmetric shell structure, the density follows the relation $N(R,\theta) = A/R^2 \times (s_0|\sin\theta|^m + 1-s_0)$, where parameters $m$ and $s_0$ capture the level of angular asymmetries. $s_0 = 0$ indicates that the axisymmetric shell is reduced to a spherical shell, and the range of $s_0$ is from $0$ to $1$. The parameter $m > 0$ represents the degree of dust-gathering in the direction of the equator. With the definition of the number density of dust grains $N(R)$, the optical depth $\tau$ in the radial direction is expressed as $\int N(R)\sigma_{\text{ext}}dR$. Thus, three parameters ($R_{\text{in}}, R_{\text{wid}}, \tau$) are needed to define the geometric properties of a spherical shell. Four parameters are needed for an axisymmetric disk: ($R_{\text{in}}, R_{\text{wid}},\tau, \theta_{\text{disk}}$). Five parameters are needed for an axisymmetric shell: ($R_{\text{in}}, R_{\text{wid}},\tau, m, s_0$). Notice that the optical depth of the axisymmetric shell is defined in the direction with the maximum number density of dust grains. The angle to the observer $\theta_{\text{obs}}$ is needed as an additional parameter for the axisymmetric shell and disk structures. 

The likely values for the parameters are poorly known. SN~2002ic-like supernova represents an extreme case where the progenitor has lost a rather large amount of matter shortly before the SN explosion \citep{Hamuy2003Natur.424..651H,Wang2004ApJ...604L..53W,Aldering:2006ApJ...650..510A}. Spectropolarimetry shows that the interaction between the SN ejecta and the CSM is highly asymmetric \citep{Wang2004ApJ...604L..53W}. In the recurrent nova scenario developed by \cite{2012ApJ...761..182M} for these supernovae, a diffusing medium-velocity ($\sim 10-100$ km s$^{-1}$) CSM was ejected shortly before the supernova explosions. Spectroscopically normal SNe~Ia may have CSM at significantly larger distances but this has so far escaped any observational detection. In this study, the dusty CSM is restricted to being at distances around $10^{17}$ cm following the work of \cite{Wang:2019ApJ...882..120W}. 

\begin{table}
\begin{center}
\begin{tabular}{cccc}
\hline
\  & Parameter Range &  Numbers of Grids & \  \\ 
\hline
$R_{\text{in}}$ & $[20, 200]$ & $19$ & S, D, A \\
$R_{\text{wid}}$ & $[20, 200]$ & $19$ & S, D, A \\
$\tau$ & $[0.1, 1.0]$ & $10$ & D \\
$\tau$ & $[0.05, 0.25]$ & $5$ & S, A \\
$\theta_{\text{disk}}$ & $[6^{\circ},30^{\circ}]$ & $5$ & D \\
$(m,s_0)$    &   $[(0.5,0.1), (5.0,1.0)]$ & $10$  &  A \\
$\theta_{\text{obs}}$ & $[10^{\circ},90^{\circ}]$ & $9$ & D, A  \\
\hline
\end{tabular}
\caption{CSM parameter ranges and grid numbers of the three structures. The grids is uniformly distributed in the parameter ranges. The terms 'S', 'D', 'A' represent the spheric shell, axisymmetric disk, and axisymmetric shell respectively. Here, $R_{\text{in}}$ and $R_{\text{wid}}$ are measured in light-days. }
\label{table_para_setting}
\end{center}
\end{table}

\begin{table}
\begin{center}
\begin{tabular}{ccccccc}
\hline
\  & $R_{\text{in}} \text{(lt-day)}$ & $R_{\text{wid}} \text{(lt-day)} $ & $\tau$ & $\theta_{\text{disk}}$ & $(m,s_0)$ & $\theta_{\text{obs}}$ \\
\hline
S1 & $40$ & $40$ & $0.15$ & \   & \    & \    \\
S2 & $50$ & $20$ & $0.15$ & \   & \    & \    \\
S3 & $20$ & $140$ & $0.2$ & \   & \    & \    \\
D1 & $40$ & $40$ & $0.5$ & $18^{\circ}$ & \    & $30^{\circ}$    \\
D2 & $20$ & $150$ & $0.7$ & $12^{\circ}$ & \    & $10^{\circ}$    \\
D3 & $30$ & $70$ & $0.3$ & $24^{\circ}$ & \    & $20^{\circ}$    \\
D4 & $140$ & $20$ & $0.8$ & $24^{\circ}$ & \    & $60^{\circ}$    \\
D5 & $140$ & $110$ & $0.9$ & $30^{\circ}$ & \    & $90^{\circ}$    \\
A1 & $40$ & $40$ & $0.15$ &  \  & $(2.5, 0.5)$  & $30^{\circ}$    \\
A2 & $20$ & $120$ & $0.25$ &  \  & $(4.5, 0.9)$  & $10^{\circ}$    \\
A3 & $30$ & $90$ & $0.15$ &  \  & $(0.5, 0.1)$  & $90^{\circ}$    \\
A4 & $40$ & $30$ & $0.2$ &  \  & $(3.5, 0.7)$  & $90^{\circ}$    \\
\hline
\end{tabular}
\caption{S*, D*, and A* represent the chosen CSM parameters of the shell, disk, and axisymmetric shell structures, respectively. S1, D1, and A1 are the corresponding reference CSM parameters. The unit lt-day denotes light-day.}
\label{table_para_choose}
\end{center}
\end{table}

\section{Results} 
\label{sec_result}

\subsection{Kernel of Intensity} 

\begin{figure}
\centering
\includegraphics[width = 0.9 \linewidth]{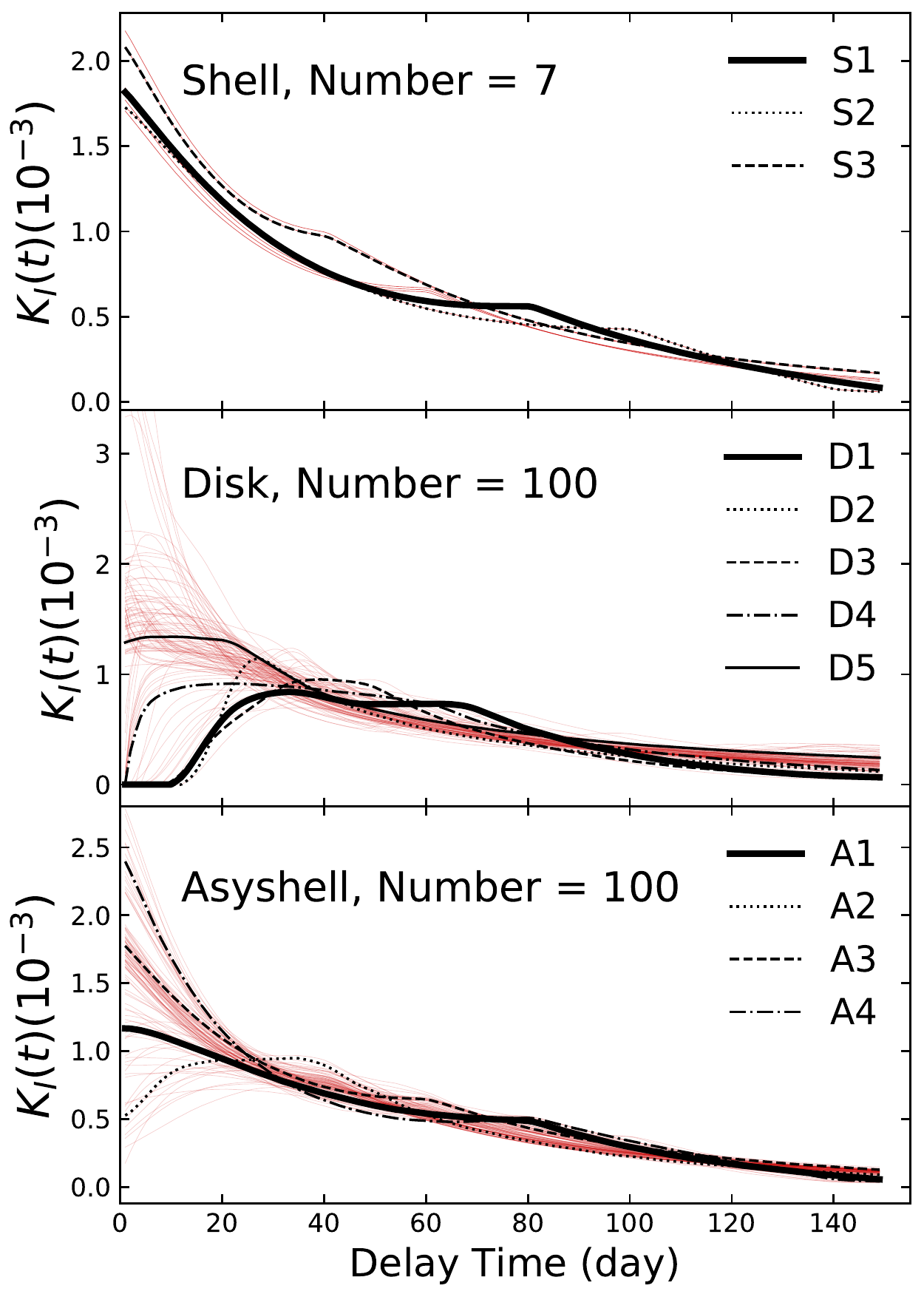}
\caption{Upper panel: kernel functions $K_I(t)$ that are similar to the reference case S1 for the spherical shell model. The black solid, dotted, and dashed lines are for models S1, S2, and S3, respectively. The other similar models are shown as thin red lines. Middle panel: kernel functions that are similar to the reference case D1. The black thick solid, dotted, dashed, dashed-dotted, and thin solid lines show models D1, D2, D3, D4, and D5. The red lines show 100 models randomly chosen from the $881$ $K_I(t)$ models that are similar to D1. Lower panel: kernel functions that are similar to the reference case A1. The black thick solid, dotted, dashed, and dashed-dotted lines show models A1, A2, A3, and A4, respectively. The thin red lines show $100$ models randomly selected models from the $564$ $K_I(t)$ models that are similar to D1. The details of the models shown in black lines can be found in Table~\ref{table_para_choose}. } 
\label{fig_kernel_i00}
\end{figure} 

The kernel function $K_I(t)$ is the distribution of scattered photons as a function of the delay time $t$ for a $\delta$-function impulse of input light. This distribution is affected by physical properties of the CSM and its geometry. However, a variety of CSM parameters may produce very similar kernel functions $K_I(t)$ and this introduces a considerable amount of degeneracy, which makes it difficult to disentangle the various effects involved. For an axisymmetric disk or shell, observers at smaller $\theta_{\text{obs}}$ will detect a broader range of time delays, similar to the effect caused by a larger $R_{\text{in}}$. Larger $\theta_{\text{disk}}$ for an axisymmetric disk, smaller $(m, s_0)$ for an axisymmetric shell, and larger values of $\tau$ all lead to a larger number of scattered photons. 

To understand such degeneracy, we calculated the kernel functions for parameter grids that cover a broad range of the geometric distribution of the CSM. Table~\ref{table_para_setting} shows the configuration of the CSM parameter grids. The total  number of the parameter grids is $1805$ ($19 \times 19 \times 5$) for the spherical shell, and $18,050$ ($19 \times 19 \times 5 \times 10$) for the axisymmetric disk and axisymmetric shell models. For each simulation of the axisymmetric disk or axisymmetric shell, nine observing angles uniformly distributed from $10^{\circ}$ to $90^{\circ}$ were calculated. The degeneracy of CSM parameters is complicated. For illustrative purposes only, we defined three reference sets of CSM parameters S1, D1, and A1 for the spherical shell, axisymmetric disk, and axisymmetric shell, respectively, to examine the parameter degeneracy. The parameters that define reference sets are shown in Table~\ref{table_para_choose}. The CSM parameters of these characteristic sets are consistent with the fitting results in \citet{Wang:2019ApJ...882..120W} and \citet{2019ApJ...882...30L}, where the likely distances from the CSM around a few high-velocity SNe~Ia were found to be approximately $(1-2)\times10^{17}$ cm. The optical depths were found to be around $0.7$ for the axisymmetric disk model and $0.15$ for the spherical shell and axisymmetric shell models. 

Scattered light close to the optical maximum is mixed with the bright SN light and is hard to detect photometrically. Late-time data are more useful in quantitative diagnostics of the circumstellar dust. The degeneracy of the kernel function after maximum light can be evaluated quantitatively by defining two measures: the average of $K_I(t)$ from $20$ days to $100$ days $K_{I\text{mean}} = \sum_{i=20}^{i=100} K_I(i) / 81$, and the ratio of the intensities at $100$ days and $20$ days $K_{I\text{ratio}} = K_I(100) / K_I(20) $. The similarity of the kernel function $K_I(t)$ is defined by the following criteria:  $|K_{I\text{mean}} - K_{I\text{mean}}^0| / K_{I\text{mean}}^0 < 0.1$ and $|K_{I\text{ratio}} - K_{I\text{ratio}}^0| / K_{I\text{ratio}}^0 < 0.1$, where $K_{I\text{mean}}^0 $ and $K_{I\text{ratio}}^0$ correspond to the values for the reference models S1, D1, or A1. 

With the above criteria, seven sets of spherical shell models share similar late-time kernel distributions to the reference model S1, while for the reference cases D1 and A1, $881$ and $564$ sets show similar late-time kernel functions, respectively. For the three geometric models of the CSM, the fraction of late-time kernel functions that are similar to their corresponding reference models is less than $1\%$ of the total number of models.  Figure~\ref{fig_kernel_i00} shows all of the kernel functions similar to S1 at late time for the spherical shell model in the top panel and $100$ models randomly selected from similar models for the axisymmetric disk and axisymmetric shell models (middle and bottom panels). For comparison, several characteristic cases are highlighted for the spherical shell model (S1, S2, and S3), axisymmetric disk model (D1, D2, D3, D4, and D5), and axisymmetric shell (A1, A2, A3, and A4). The individual CSM parameters for these characteristic cases are listed in Table~\ref{table_para_choose}. The degeneracy is  obvious; e.g., for the axisymmetric shell model, the large $\tau$ and $(m,s_0)$ values in case A2 and the small corresponding values in case A3 result in a similar kernel function $K_I(t)$. As we just discussed, with this kernel function degeneracy, the CSM parameters cannot be determined by fitting the light-curve data only.

\begin{figure}
\centering
\includegraphics[width = 0.9 \linewidth]{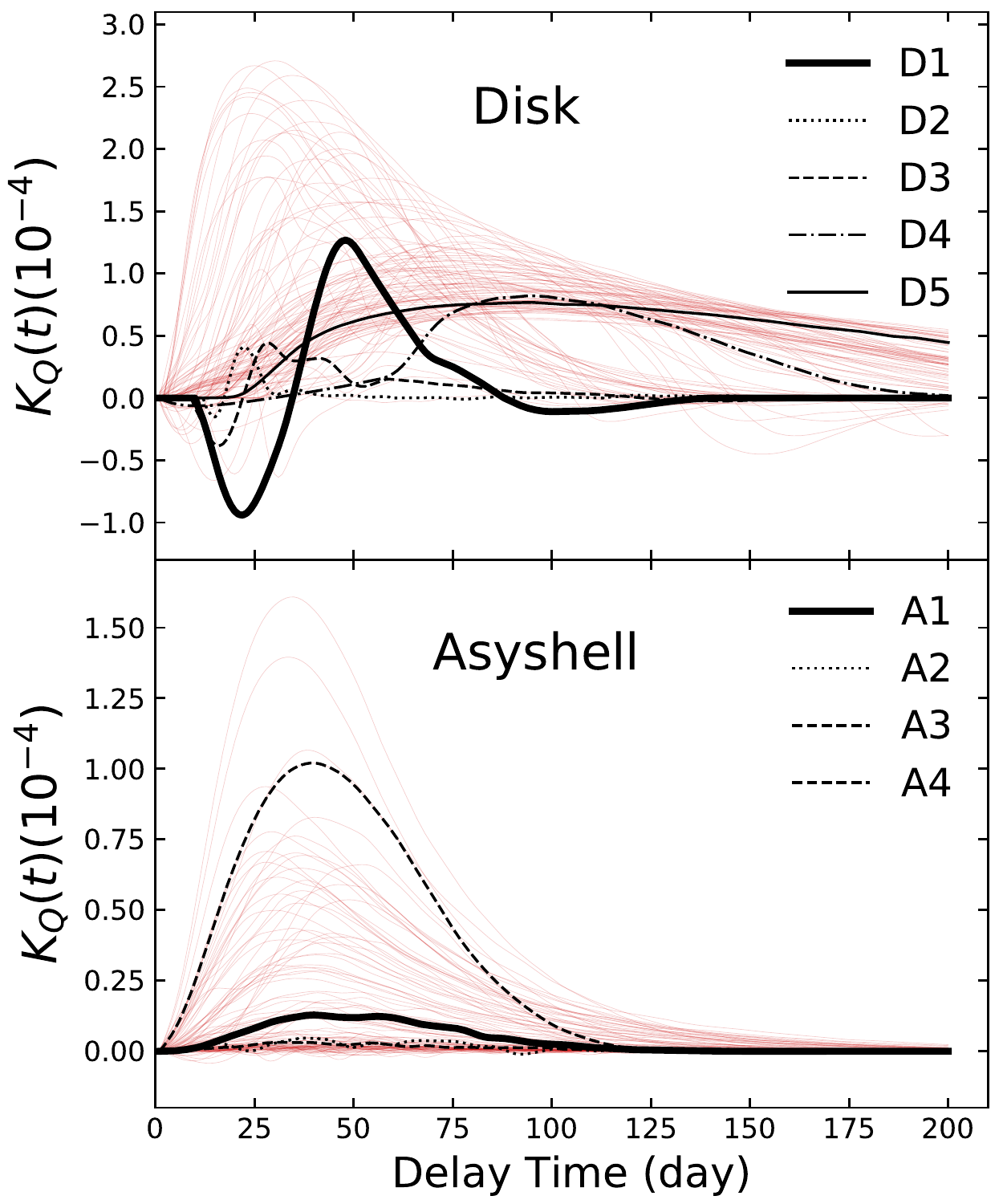}
\caption{Same as Figure~\ref{fig_kernel_i00}, but for the kernel function of Stokes parameter $Q$ just for the disk model (upper panel) and axisymmetric shell model (lower panel). They have the same CSM parameters as the 100 randomly selected cases for both models.}
\label{fig_kernel_q00}
\end{figure}

\begin{figure*}[ht]
\centering
\includegraphics[width = 0.95 \linewidth]{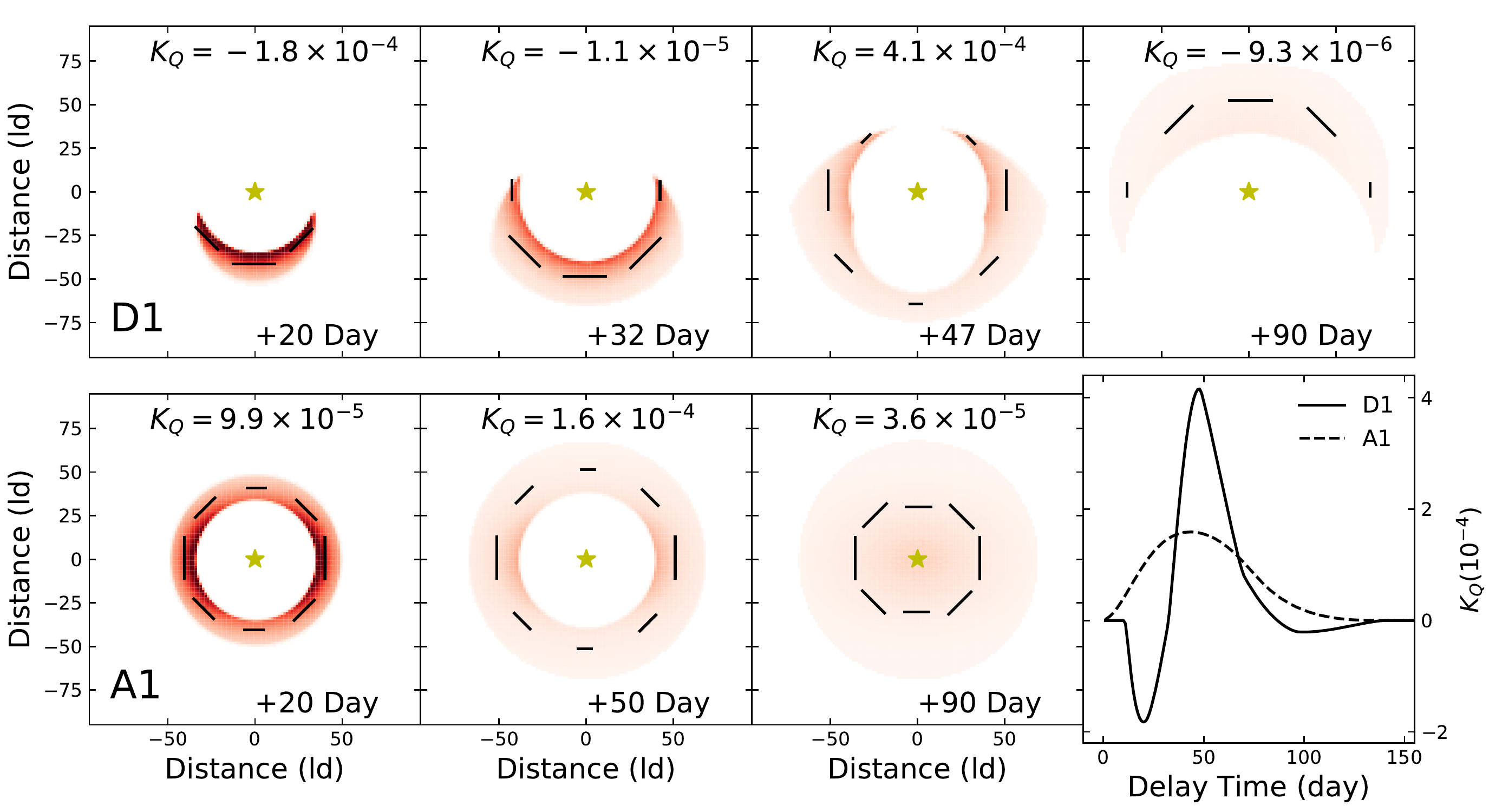}
\caption{Upper panels: the intersecting area between the parabolic surface and the CSM projecting to the line of sight (red region) and $Q-U$ vectors (short black lines) of four epochs ($+20$ days, $+32$ days, $+47$ days, and $+90$ days) for the disk model D1. Lower panels: three epochs ($+20$ days, $+50$ days, and $+90$ days) for the axisymmetric shell model A1; far right lower panel: the corresponding $K_Q(t)$ curves of the disk and axisymmetric shell models. The observing angle $\theta_{\text{obs}}$ for both the models A1 and D1 is $30^{\circ}$ and the other parameter values are shown in Table~\ref{table_para_choose}.}
\label{fig_qu_single}
\end{figure*}

\subsection{Kernel of the Stokes Parameter Q}

Polarization can be a powerful diagnostic tool if dust scattering is indeed important. For the spherical shell, the polarization of the scattered photons cancels out, and there would be no net polarization. On the other hand, the scattered light from the axisymmetric disk or axisymmetric shell may be highly polarized. Without loss of generality, we will assume that the axis of symmetry of the disk is pointing north, the Stokes parameter $U$ of the axisymmetric disk and axisymmetric shell is zero, and only the Stokes parameter $Q$ is nonzero, with the degree of polarization $P = |Q| / I$.    

The degree of polarization is the most significant when the target is viewed edge-on ($\theta_{\text{obs}} = 90^{\circ}$), and is zero when it is viewed face-on ($\theta_{\text{obs}} = 0^{\circ}$). In addition to the geometric distribution, polarization also depends on the optical cross section and albedo of the dust grains. Again, the calculation of the polarization can be calculated by first calculating the kernel function $K_Q(t)$ for the Stokes parameter $Q$, by assuming the light source is a $\delta$-function. 

The different CSM parameters that generate very similar kernel functions of the intensity (Figure~\ref{fig_kernel_i00}) now generate dramatically different kernel functions for the Stokes parameter $Q$. This demonstrates that the combination of $K_I(t)$ and $K_Q(t)$ can distinguish the different dust geometries and thus break the degeneracy. Figure~\ref{fig_kernel_q00} shows the  kernel function $K_Q(t)$ of the $100$ cases shown in Figure~\ref{fig_kernel_i00} for the axisymmetric disk and axisymmetric shell models. The polarization curves show a broad range of behaviors, which makes them very powerful in establishing the presence of CS dust and constraining their geometric structures. As an example, $K_Q(t)$ of D3 is smaller than that of D5 owing to a smaller $\theta_{\text{obs}}$, and the time evolution of the degree of polarization is sensitive to the geometric size and location of the dust. With the same $\theta_{\text{obs}}$ of $90^{\circ}$, A3 and A4 have distinctively different $K_Q(t)$ owing to their different values of $(m, s_0)$.

\begin{figure*}[t]
\centering
\includegraphics[width = 0.95 \linewidth]{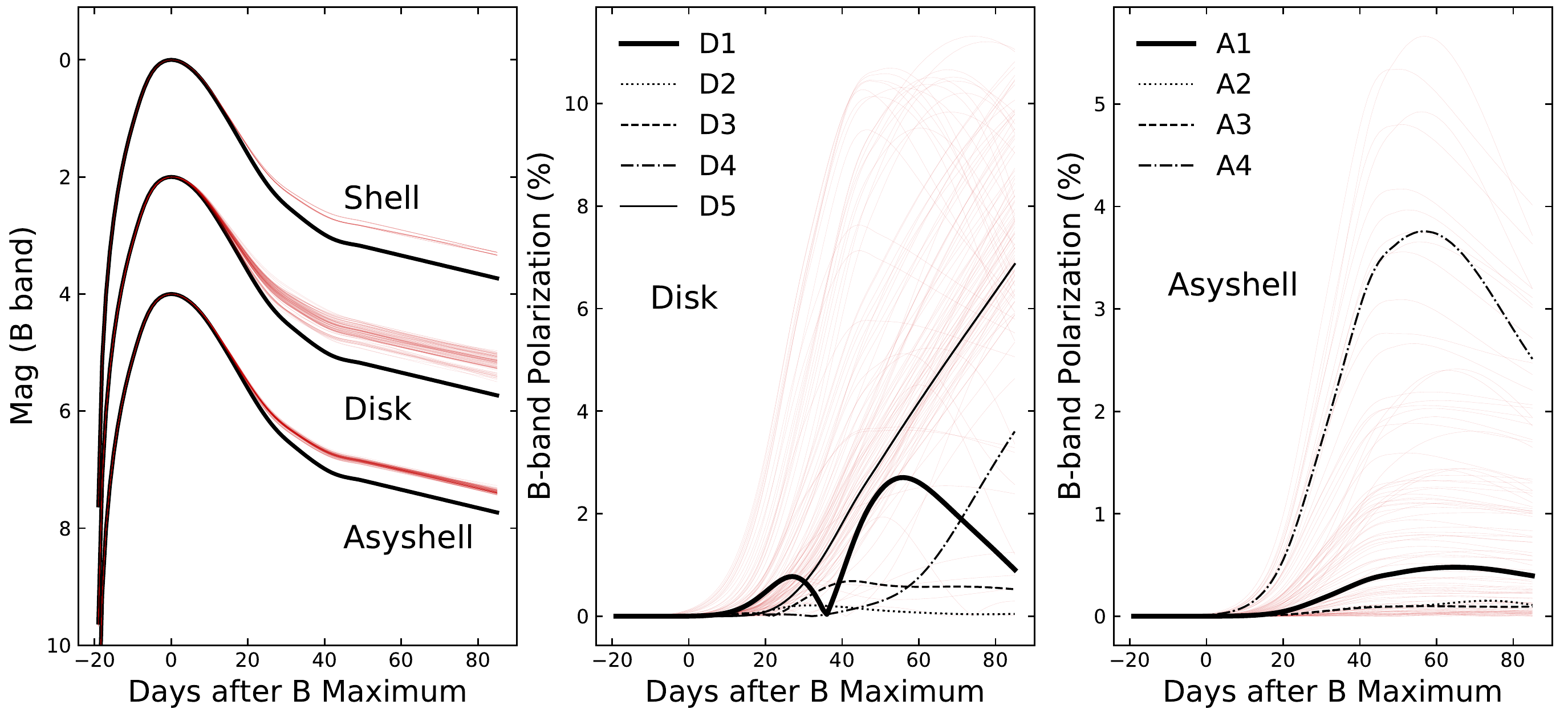}
\caption{Left panel: the predicted $B$-band light curves (red lines) with the same CSM parameter values as in Figure~\ref{fig_kernel_i00} vs. the $B$-band template of SNe~Ia (black lines). All the light curves are scaled to their maximum light.} Middle and right panels: the predicted polarization curves of a disk and an axisymmetric shell, respectively.
\label{fig_lcs_polar}
\end{figure*}

\subsection{Q-U Distribution for Reference Cases D1 and A1}

Light echoes can be used as a tomographic method that can effectively probe the 3D geometry of the scattering material. This tool becomes even more powerful with the inclusion of polarimetry. It is interesting to note that the two  models D1 and A1 have very different $K_Q(t)$ curves, but with geometric structures that are rather similar (Figure~\ref{fig_kernel_q00}). The differences can be examined by calculating the surface brightness of the scattered light and the 2D $Q-U$ distributions for the reference cases A1 and D1. For the purpose of making the figures, we assumed the single-scattering approximation. The results are shown in Figure~\ref{fig_qu_single}.

The axisymmetry ensures that the integrated Stokes parameter $U$ is 0, therefore only the $Q$ component of the Stokes parameter needs to be considered. For the axisymmetric shell structure, $Q$ from the equatorial region is always larger than $Q$ from the two polar directions. Thus, $K_{Q}(t)$ is positive with any $\theta_{\text{obs}}$ or any values of CSM parameters for the axisymmetric shell model A1. This means that $K_Q(t)$ never changes signs, as shown in the bottom panel of Figure~\ref{fig_kernel_q00}. While for the disk model, the polarization may be dominated by scattering from either the equatorial or the polar regions depending on the epoch of observations. This causes $K_Q(t)$ to change sign with time, as shown in Figure~\ref{fig_qu_single} and  Figure~\ref{fig_kernel_q00}. Note that the degrees of polarization are slightly different in Figure~\ref{fig_qu_single} and  Figure~\ref{fig_kernel_q00} for models D1 and A1. This is because multiple scattering is assumed in Figure~\ref{fig_kernel_q00} but the single-scattering approximation is assumed in Figure~\ref{fig_qu_single} for illustrative purposes.

\section{The Scattered Light of Type Ia Supernovae}
\label{discussion}

In this section, the kernel functions are convolved with an spectral energy distribution (SED) template to predict the light curves, polarization, and spectral evolution of Type Ia supernovae. We will also apply these models to fit the  $E(B-V)$ color curves, as has been done previously in \cite{Bulla2018MNRAS.479.3663B}, but with the goal of studying the degenerate nature of the model parameters and the difficulties in uniquely constraining the CS dust geometry without a detailed time sequence of polarimetry. 

\subsection{The Light Curves and Polarization} 

The template for light curves or spectra should come from SNe~Ia without CS dust in their vicinity. Here, the spectral template is adopted from \citet{2007ApJ...663.1187H}. This template is used to derive the light curves by applying the filter transmission functions. Figure~\ref{fig_lcs_polar} shows the $B$-band light curves and polarizations for the dust models we have investigated, obtained by convolving the spectral template with the relevant kernel functions derived in the previous section. A common feature of the models with CS dust scattering is a flux excess a month or so after the maximum brightness.

As a consequence of the sensitivity of the kernel function $K_Q(t)$ to the dust distribution geometry, the predicted polarization curves are dramatically different for different model parameters. This makes polarimetry a promising tool for constraining the dust distribution around SNe~Ia. We note that the majority of disk models predict large degrees of polarization that are observable for nearby supernovae. For the parameters we have adopted, the axisymmetric shells predict  polarization degree that are in general lower than $1.0\%$. In both the axisymmetric shell and disk cases, the degree of polarization peaks at around 50 days past optical maximum, and for the axisymmetric disk model the degree of polarization can be as large a few percent. A time sequence of polarimetry at $\sim$ 2 months can be used to test these models and establish or disprove the existence of CS dust around SNe~Ia. No polarization evolution at such late phases has been acquired for any SN~Ia so far. 

\subsection{Constraining the Distance from Multiple-epoch Polarization}

\begin{figure*}
\centering
\includegraphics[width=0.95\linewidth]{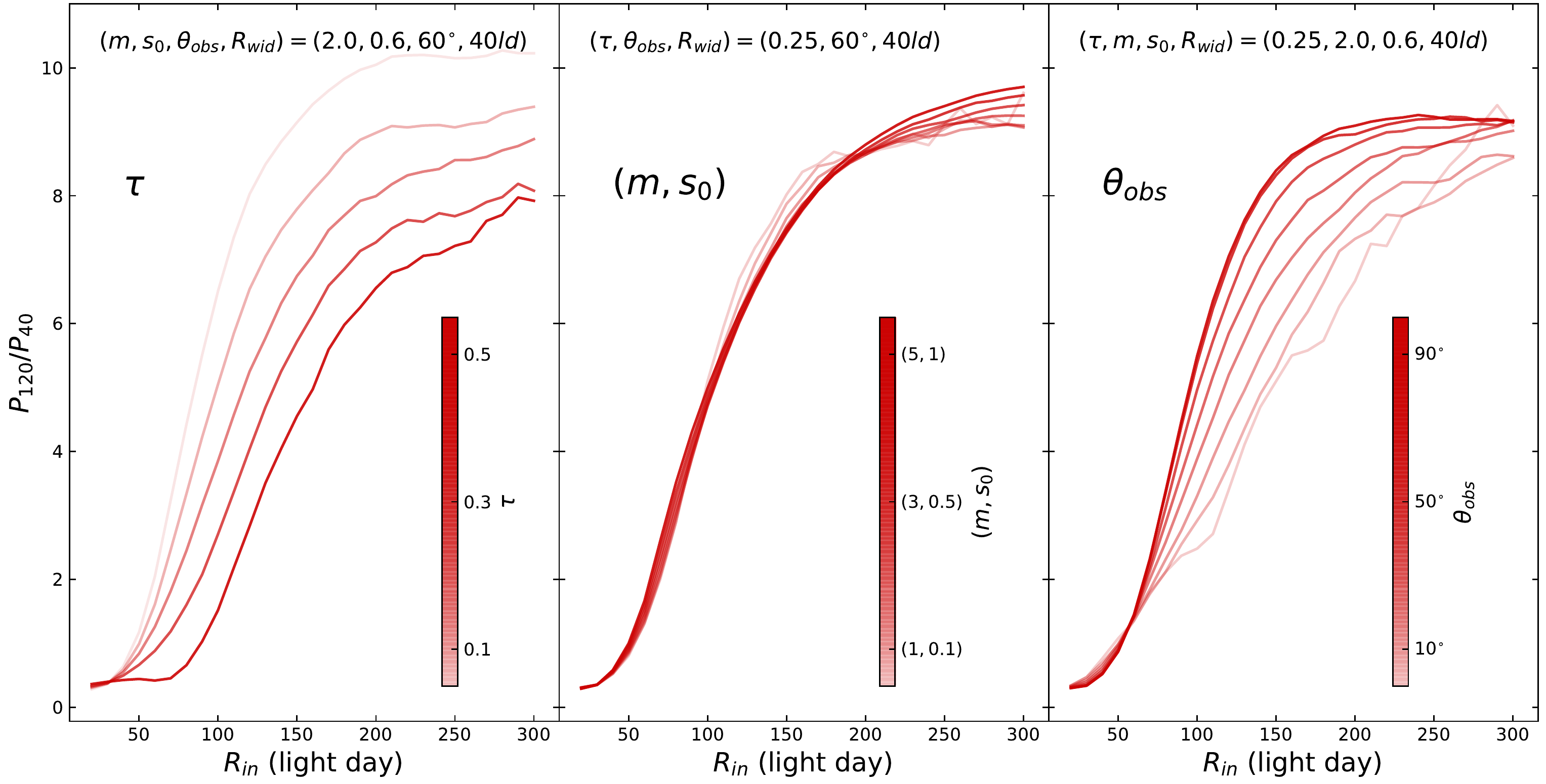}
\caption{The polarization ratio $P_{120}/P_{40}$ is approximately a monotonic function of $R_{\text{in}}$ as shown in each panel. The depth of the color represents the changing values of parameter $\tau$ (left panel), $(m, s_0)$ (middle panel), and $\theta_{\text{obs}}$ (right panel). For each configuration, the thickness of the CS dust in the radial direction $R_{\text{wid}}$ is set to $40\ \text{lt-day}$. For each panel, the values of the other relevant parameters are shown at the top of each panel.}
\label{fig_ppp}
\end{figure*}

The results above suggest that the combined observation of the photometry and polarization is a promising probe for constraining CSM features, which is based on the results that similar light curves may be related to a variety of CSM parameters while the corresponding polarization curves may help to break this degeneracy. In this section, we show that polarimetry is a crucial probe for constraining the CSM around SNe~Ia. 

Take the axisymmetric shell models as examples: the degree of polarization is sensitive to $\tau$, $(m,s_0)$, and $\theta_{\text{obs}}$. The values of these two parameters affect the overall levels of  polarization. On the other hand, the inner or outer boundaries of CSM are sensitive to the time evolution of the degree of polarization. These properties can be employed to constrain the location of the CS dust.

In order to quantify the effect of the CSM boundary on the degree of polarization, we calculated the ratio of the degrees of polarization at $+120$ and $+40$ days, $P_{120}/P_{40}$. If the distance of CSM is significantly smaller than 40 lt-day ($\sim1\times10^{17}$ cm), the typical delay time of scattered photons is small and the degree of polarization at $+40$ days is usually larger than that at $+120$ days. But if the distance of CSM is mostly around 120 lt-day ($\sim3\times10^{17}$ cm), the polarization ratio may just be the opposite. 

Figure~\ref{fig_ppp} shows the relationship between the polarization ratio $P_{120}/P_{40}$ and the inner boundary of CSM. It can be clearly seen that for different values of $\tau$, $(m, s_0)$, and $\theta_{\text{obs}}$, an approximately monotonic relationship can be established between the polarization ratio $P_{120}/P_{40}$ and the location of the inner boundary $R_{\text{in}}$. For the polarization ratio shown in Figure~\ref{fig_ppp}, $R_{\text{wid}}$ is set to $40\ \text{lt-day}$ in all simulations. As expected, Figure~\ref{fig_ppp} shows also that the polarization ratio can be dependent on the optical depth and the direction of the observations, but the sensitivity relative to the parameters describing the level of asymmetry is rather weak.

\begin{figure}
\centering
\includegraphics[width = 0.9 \linewidth]{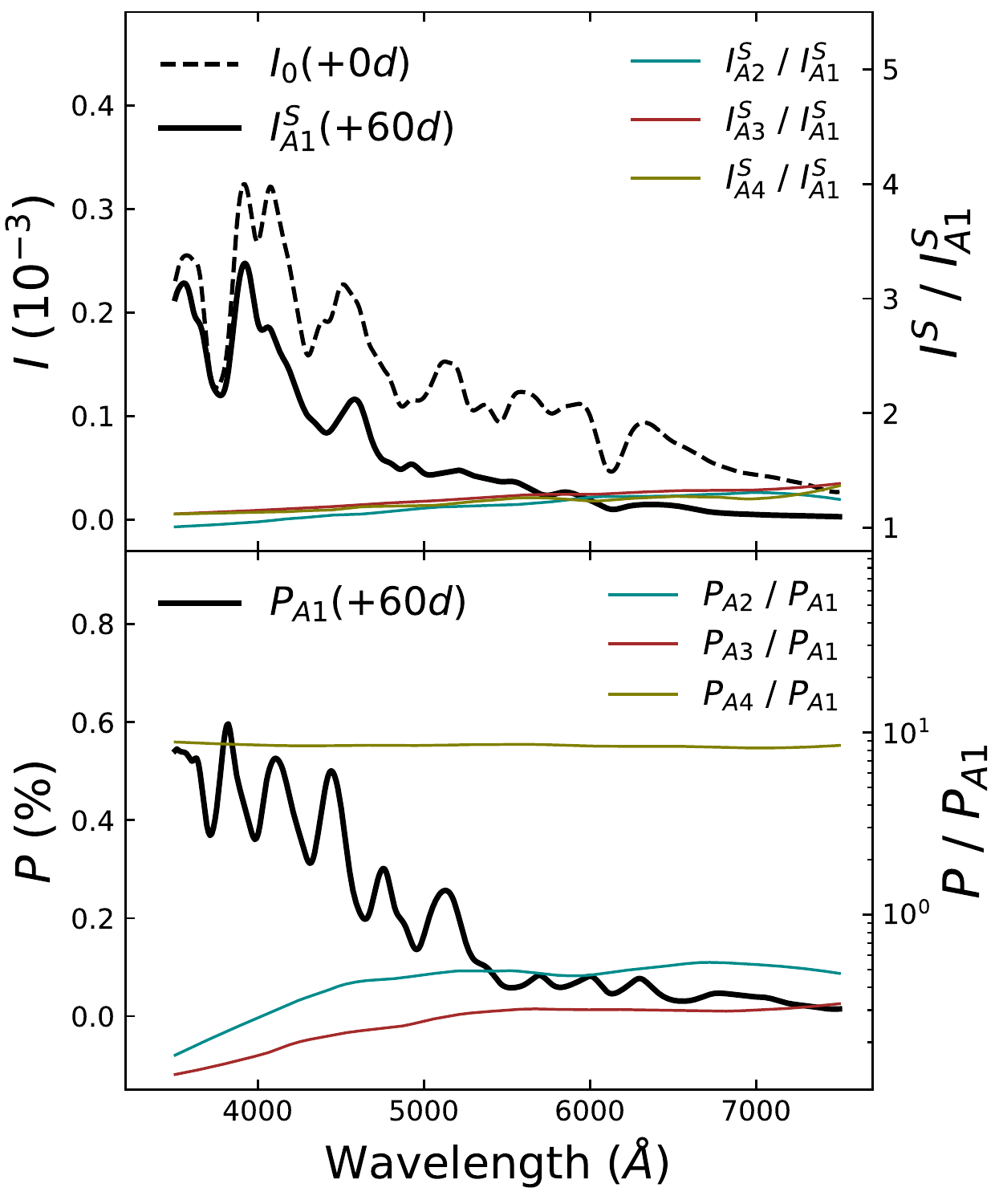}
\caption{Upper panel: the dashed black line shows the spectral template ($I_0$) of SNe~Ia at maximum light reduced to an arbitrary scale for clarity. The black line shows the scattered spectrum of the CSM model A1 at $+60$ days after peak brightness. The three colored solid lines show the ratios of the scattered intensities ($I^S / I^S_{\text{A1}}$) of the CSM models A2, A3, and A4 and model A1. Lower panel: the black line is the polarization spectrum of the CSM model A1. The three colored solid lines are the ratios of the polarizations ($P / P_{\text{A1}}$) of the CSM models A2, A3, or A4 and the model A1. The values of the CSM parameters of A1, A2, A3, and A4 are listed in Table~\ref{table_para_choose}.}
\label{fig_spectra}
\end{figure}

\begin{figure}
\centering
\includegraphics[width = 0.9 \linewidth]{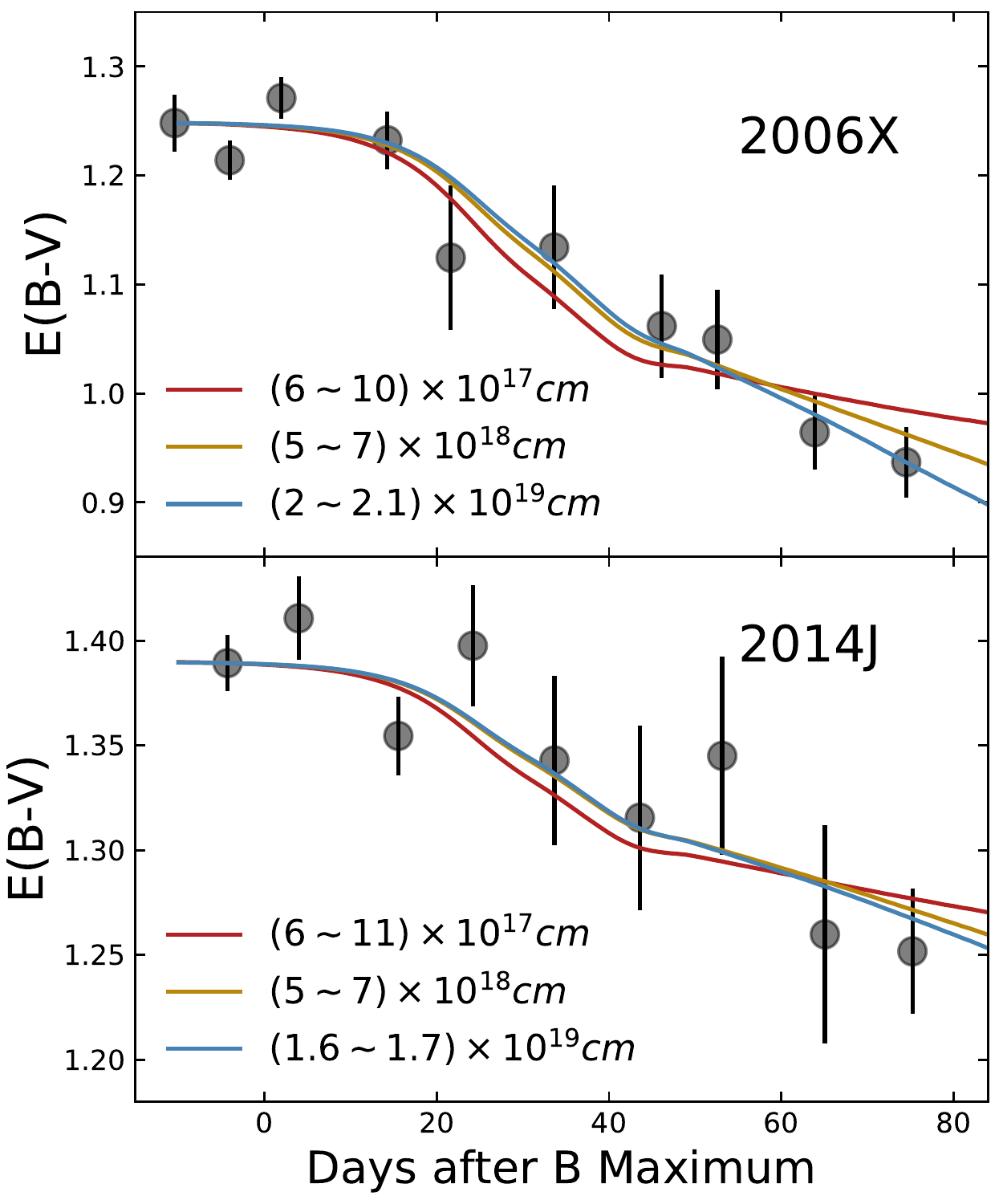}
\caption{The circles are the data of $E(B-V)$ curves of SN~2006X (top panel) and SN~2014J (bottom panel) from \citet{Bulla2018MNRAS.479.3663B}. The three lines in each panel represent three shell models with different distances from the center, and they have been shifted to match the $E(B-V)$ along the line of sight for SN~2006X and SN~2014J. This shows that the color curves alone cannot provide strong constraints on the location of the CS dust.}
\label{fig_color_06x}
\end{figure}

\subsection{The Spectra of Type Ia Supernovae with an Axisymmetric Dusty Circumstellar Shell} 

The spectroscopic and spectropolarimetric evolution of SNe~Ia can be affected by the presence of asymmetric dusty CSM. As an example, Figure~\ref{fig_spectra} shows the spectrum of the scattered light and the corresponding spectropolarimetry at day 60 after optical maximum of a typical SN~Ia, for the parameter sets A1, A2, A3, and A4 (see Table~\ref{table_para_setting} for details). In the top panel, we show a spectrum of the scattered light at day 60 for the reference case A1 (black solid line), which is quite similar to the adopted spectral template (black dashed line) at optical maximum. This similarity suggests the scattered photons are dominated by those from the peak brightness. Among the models we have explored, the CS dust geometry has only a weak effect on the spectral features of the scattered light. For example, the ratios of scattered spectra of A2, A3, and A4 to that of the reference case A1, shown as the colored lines in Figure~\ref{fig_spectra}, exhibit no strong spectral modulation in the wavelength range from $350$ to $750$ nm. Similar behavior can be seen in the degree of polarization shown in the bottom panel, although the degrees of polarization are significantly different for different models. In general, the fitting of spectropolarimetry can place tighter constraints on the dust properties, such as the chemical composition and the size distribution of the dust grains, but a time sequence of broadband polarimetry is sufficient to constrain the geometric shape of the CS dust. Densely time-sampled spectropolarimetry (e.g., more than two observations in late phases) can be difficult when considering observational cost but is fortunately not needed. 

\subsection{The E(B-V) Curves of SN~2006X and SN~2014J and Their CS Dust}

SN~2006X \citep{2008ApJ...675..626W} and SN~2014J \citep{2015ApJ...798...39M,2016MNRAS.457.1000S,2017ApJ...834...60Y} are two highly reddened nearby supernovae. They can serve as good examples to study the location of the dust along the lines of sight to the SNe.

Dust scattering is color-sensitive and, if present, can alter the evolution of the  color excess $E(B-V)$. \cite{Bulla2018MNRAS.479.3663B} adopt a thin shell geometry for the CS or interstellar dust to model the color excess $E(B-V)$ curves of SNe~Ia to place constraints on the location of the dust. A single spherical shell is used to simultaneously fit the large values of $E(B-V)$ and its time evolution. Therefore, the optical depth of the shell is fixed by the total reddening. In their models, the radius of the inner boundary is set to 0.95 times of the radius of the outer boundary. The dust distribution is uniform in the shell. Their models assume the Henyey$-$Greenstein dust scattering phase function (\citealt{1941ApJ....93...70H}) and Milky Way-like dust grains. The radii of the dusty shells for SN~2006X and SN~2014J are found to be $44.6\ \text{pc}$ (or $\sim10^{20}$ cm) and $17.3\ \text{pc}$ (or  $\sim5\times10^{19}$ cm), respectively, according to these models, thus placing the dust grains at distances that are typically beyond those for CSM. These distances are also much larger than the distances of the putative CSM derived by \cite{Wang:2019ApJ...882..120W} based on the evolution of the narrow Na ID lines.

In reality, the distribution of the dust responsible for the heavily reddened SNe such as SN~2006X and SN~2014J may be rather complicated. The extinction may come from the interstellar dust across the host galaxy along the line of sight (e.g., the spiral arm area), the dusty interstellar environment close to SNe~Ia (e.g., a few parsecs as shown in \citet{Bulla2018MNRAS.479.3663B}), or from CS dust. In this paper, we assume that the extinction of highly reddening SNe~2006X and 2014J comes from the interstellar dust across the host galaxy and the CS dust around SNe~Ia. Thus, the interstellar dust is less likely to be the cause of time-varying reddening, and only the time evolution of the $E(B-V)$ may likely reveal the CS dust. Both the scenarios shown in \citet{Bulla2018MNRAS.479.3663B} and in our work can explain the time evolution of $E(B-V)$ reasonably, but polarimetry (as discussed in our work) and thermal emission from dust in the CSM are efficient probes to distinguish them.   

As we have shown already, there is a considerable amount of degeneracy among the model parameters. To compare with the  results of \cite{Bulla2018MNRAS.479.3663B}, we consider the simple spherical shell model at three distances of $\sim10^{17}$ cm, $\sim10^{18}$ cm, and $\sim10^{19}$ cm to fit the $E(B-V)$ color curves of SN~2006X and 2014J. For SN~2006X, the optical depths are $0.3$, $2.1$, and $4.8$ for the shells at the distances of $10^{17}$ cm, $10^{18}$ cm, and $10^{19}$ cm, respectively. For SN~2014J, the optical depths are $0.12$, $0.97$, and $1.5$ at these three distances. The source of the observed $E(B-V)$ curves is the compilations of  \cite{Bulla2018MNRAS.479.3663B}, and the original sources of the data are from \citet{2008ApJ...675..626W} for SN~2006X and \citet{2015MNRAS.453.3300A} for SN~2014J. The results are shown in Figure~\ref{fig_color_06x}. All three shell models can fit the time evolution of $E(B-V)$ satisfactorily, confirming the degenerate nature of model parameters. A similar result could be acquired in the opposite way of fitting the CSM distance by fixing the optical depth $\tau$. For instance, if we fix $\tau$ with the values of $0.3$, $2.0$, and $5.0$, the corresponding values of $R_{\text{in}}$ for SN~2006X would be about $10^{17}$ cm, $10^{18}$ cm, and $10^{19}$ cm by fitting its photometric data. Meanwhile, our result is consistent with that in \citet{Bulla2018MNRAS.479.3663B} if we fix $\tau$ with some relatively large value. For instance, the shell distance for SN~2014J in \citet{Bulla2018MNRAS.479.3663B} is about $5\times10^{19}$ cm, while the distance in our work is around $1.6\times10^{19}$ cm. These two results are consistent with each other. The slight difference might be due to the dust properties, the scattering process, or the choice of the template of the light curve adopted by our models. We thus point out that even with well observed photometric data of highly extinct SNe may not be sufficient to constrain the location of the dust in the context of light echo models. Multiepoch image polarimetry is an important complementary probe to reveal the location of dust in CSM.  

\subsection{Fitting the Distance of CSM around SN~2014J through Polarization}

\begin{figure}
\includegraphics[width=0.8\columnwidth]{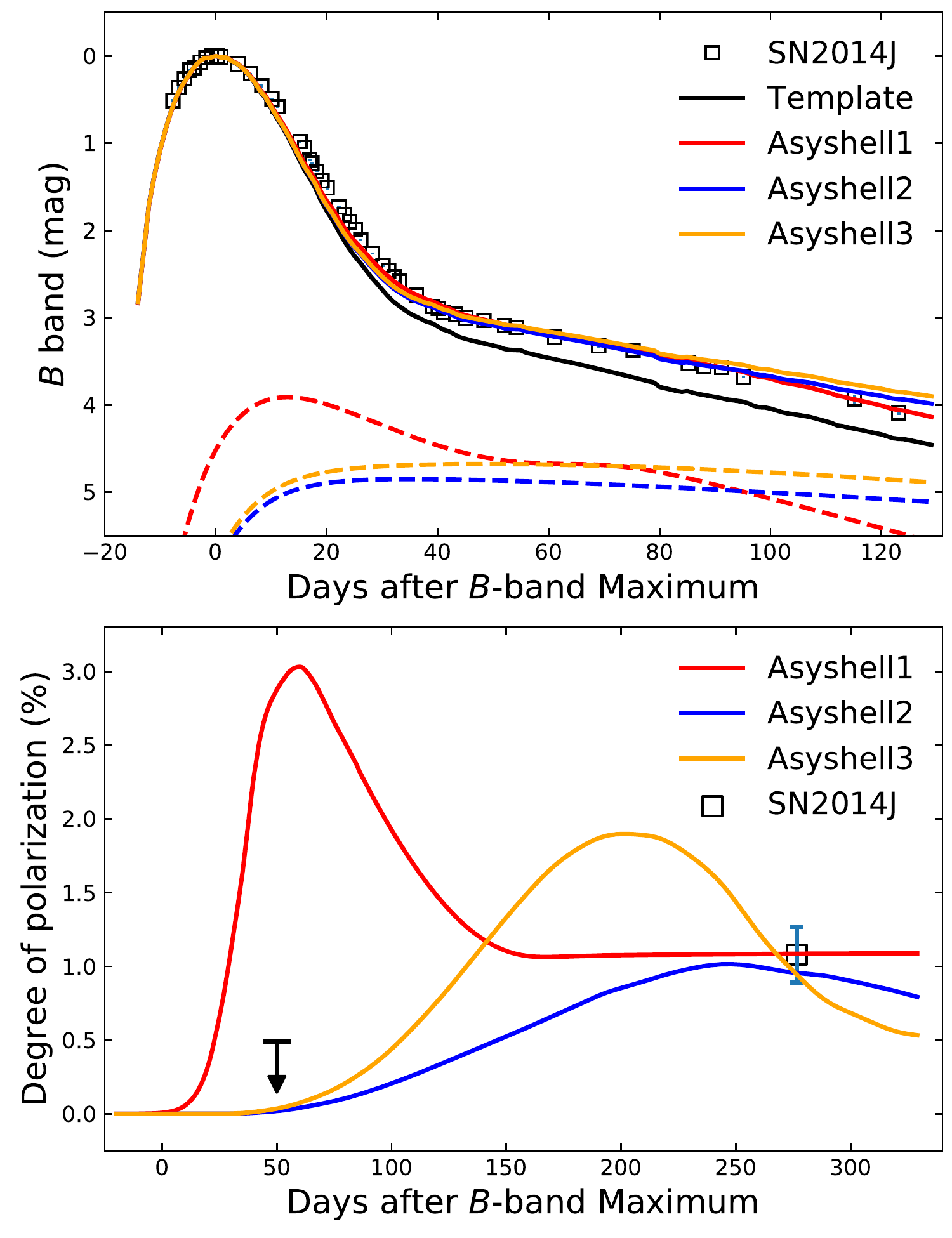}
\caption{In the upper panel, the black line is the template of $B$-band light curve. The red, blue, and orange solid lines are the fitted $B$-band light curves relating to the models Asyshell1, Asyshell2, and Asyshell3, respectively. The dashed lines are the scattered intensity. All the light curves have been scaled to the maximum light. In the lower panel, the red, blue, and orange lines are the polarization curves predicted by the models Asyshell1, Asyshell2, and Asyshell3, respectively. The parameter values of these three axisymmetric shell models are shown in Table~\ref{para_asymetric shell}.}
\label{fig_polar_14j}
\end{figure}

\begin{table}
\begin{center}
\begin{tabular*}{0.49\textwidth}{ccccccc}
\hline
\  &  $\tau$ & $R_{\text{in}}\text{(lt-day)}$  & $R_{\text{wid}}\text{(lt-day)}$    & $m$ & $s_0$ & $\theta_{\text{obs}}$ \\ 
\hline
Asyshell1 & $0.16$ & $35$ & $50$ &  $2.0$ & $0.9$ & $70^{\circ}$ \\ 
Asyshell2 & $0.24$ & $140$ & $140$ &  $3.0$ & $0.5$ & $30^{\circ}$ \\ 
Asyshell3 & $0.35$ & $200$ & $40$ &  $3.5$ & $0.7$ & $40^{\circ}$ \\ 
\hline
\end{tabular*}
\caption{The parameter values of Asyshell1, Asyshell2, and Asyshell3 models shown in Figure~\ref{fig_polar_14j}.}
\label{para_asymetric shell}
\end{center}
\end{table}

\begin{figure}
\includegraphics[width=0.9\columnwidth]{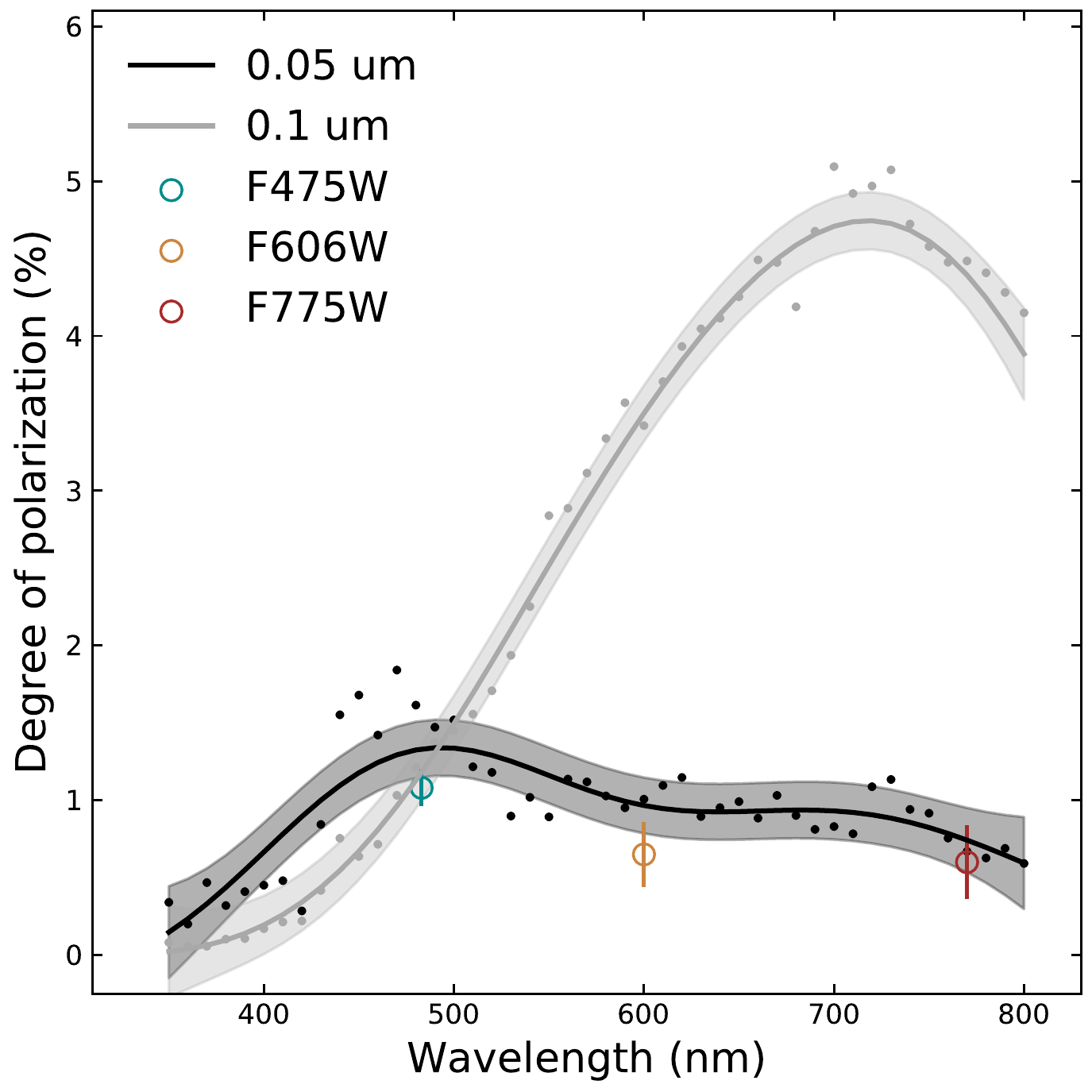}
\caption{The cyan, yellow, and brown symbols are the observed polarizations of SN~2014J during 277 days after the peak brightness for HST $F475W$, $F606W$, and $F775W$ bands from \citet{Yang2018ApJ...854...55Y}, respectively. The black and gray circles are polarization predicted by the models with dust radii of $0.05\ \mu m$ and $0.1\ \mu m$, respectively. The black and gray lines are the smoothed lines from Gaussian process fitting \citep{scikit-learn}, and the corresponding shaded regions are the 1$\sigma$ standard deviation.} 
\label{fig_polar_14j_multiband}
\end{figure}

On the one hand, the interstellar dust produces polarization through dichroic absorption, which is unlikely to show strong time evolution. On the other hand, in the scenario where the late-phase light curve of SNe~Ia includes the scattered light from interstellar dust, the scattering angle should be as small as about $5^{\circ}$, constrained by the delay time (e.g., $\sim$ 50 days) and the distance of interstellar dust (e.g., $\sim$ 10 pc). Such a small scattering angle cannot introduce significant polarization signals. Thus, we show that the time evolution of the polarization is a deterministic signature of CS dust polarization. However, there are few late-phase polarimetries (e.g., 100 days after the peak light, and see the references such as \citealt{2019MNRAS.490..578C,2022MNRAS.509.6028C}) on SNe~Ia due to the time-consuming observations. SN~2014J is one that has been observed by imaging polarimetry during such a late phase, and this provides an excellent opportunity to constrain the parameter values of CSM. As reported by \citet{Yang2018ApJ...854...55Y}, the image polarimetry shows an apparent deviation of about $1.0\%$ in the $F475W$ band of the Hubble Space Telescope (HST) at around $+277$ days after maximum light compared to the polarization at the peak brightness. This deviation is highly possible from the scattering effect of CS dust instead of interstellar dust. \citet{Yang2018ApJ...854...55Y} attributed these polarization signals to the scattering from a dusty cloud located at around $5\times 10^{17}\ \text{cm}$ from the SN. Here we apply our CS dust scattering model to study the photometry and polarimetry of SN~2014J.  

The models are constructed for the axisymmetric shell geometry. The models Asyshell2 and Asyshell3 are two sets of axisymmetric shells that can fit the photometric and polarimetric data of SN~2014J reasonably. The model parameters are shown in Table~\ref{para_asymetric shell}. The model fits to the $B$-band light curve and the polarizations are shown in Figure~\ref{fig_polar_14j}. The location of the CS dust is at distances larger than 140 lt-day (Table~\ref{para_asymetric shell}). For comparison, an axisymmetric shell with relatively close distance (Asyshell1) is also displayed, which can fit the light curves and the polarization signal up to $277$ days after maximum light precisely, but is excluded by the lack of a clear evolution in the degree of polarization at early times \citep{Kawabata2014ApJ...795L...4K,Yang2018ApJ...854...55Y}.

Obviously, the value of $P_{120}/P_{40}$ is less than $1.0$ for the Asyshell1 and is much larger than $1.0$ for both Asyshell2 and Asyshell3 models. Determining whether Asyshell2 or Asyshell3 is more reasonable for the potential distribution of CSM around SN~2014J is slightly ambiguous. Figure~\ref{fig_polar_14j} shows that Asyshell2 produces relatively small degrees of polarization at all epochs and Asyshell3 produces large polarization about $200$ days after $B$-band maximum light, though there are no observations on the polarization at the same epochs. Nevertheless, the distance of CSM around SN~2014J is about $5\times10^{17}\ \text{cm}$, which is consistent with the results in \citet{Yang2018ApJ...854...55Y}, though two different distributions (the axisymmetric shell and blob) are used respectively. The mass loss rate of the stellar wind is about $5\times 10^{-6}\ M_{\odot}\ \text{yr}^{-1}$ for the model Asyshell2, which is consistent with the observational restrictions on CSM and the progenitor of SN~2014J from $\text{H}_{\alpha}$, infrared, and X-ray signals \citep{2014ApJ...790...52M,2015A&A...577A..39L,2016ApJ...822L..16S,2017MNRAS.466.3442J}. 

\citet{Yang2018ApJ...854...55Y} also acquired the broadband polarization of SN~2014J for 277 days after maximum light in HST $F606W$ and $F775W$ bands. The corresponding degrees of polarization in $F606W$ and $F775W$ bands are about $0.65\%$ and $0.6\%$, respectively. Multiband polarimetry during such a late phase could provide an important probe to investigate the dust properties of CSM around SN~2014J, since the relationships between the scattering cross section and wavelengths are different for different dust grains. For simplicity, we considered two CS models with different dust radii. The first one is just the model Asyshell3 as shown in Table~\ref{para_asymetric shell} with the same dust radius of $0.05\ \mu m$. The other one has the same geometric distribution and same observing angle ($\theta_{\text{obs}} = 40^{\circ}$) as model Asyshell3 but a different dust radius ($0.1\ \mu m$) and different $B$-band optical depth ($\tau = 0.24$). This slightly different optical depth can induce the model with a dust radius of $0.1\ \mu m$ to match the $B$-band light curve of SN~2014J as the model Asyshell3 does shown in Figure~\ref{fig_polar_14j}. We adopted the observed spectra and light curves of SN~2011fe \citep{2016ApJ...820...67Z} to generate the spectral template covering the late phase to +300 days after the maximum light. To reduce the calculation time, the spectropolarimetry predicted by our models spans 46 wavelengths from 350 nm to 800 nm. As shown in Figure~\ref{fig_polar_14j_multiband}, we prefer the CS dust with a radius of $0.05\ \mu m$ for matching the multiband polarization signals.  

Indeed, precise polarization requires the use of large-aperture telescopes. At late times when we expect significant polarization evolution (50-300 days past the maximum light), SNe~Ia will be more than 3.5 magnitudes dimmer than at the peak light. Nonetheless, a large number of nearby SNe~Ia have been routinely found by recent SN surveys, making such a program feasible. 
 
\section{Conclusions} 
\label{conclusion}

This paper explores systematically the influence of dusty CSM on the light curves and polarizations of SNe~Ia. We first calculated the scattering kernel functions for the Stoke parameters and then constructed the light and polarization curves by convolving the spectral template of SNe~Ia with the corresponding kernel functions to obtain the model light and polarization curves. The kernel functions characterize the radiative transfer process for SNe located in a dusty environment and are obtained with the Monte Carlo method. We adopted the Mie scattering theory to calculate the dust scattering cross section, albedo, and scattering matrix based on the refractive index and the specific size distribution of silicate dust. We simulated a large number of geometric model grids to study the similarities among the kernel functions of intensity between $+20$ and $+100$ days (Figure~\ref{fig_kernel_i00}). Our study shows that the kernel functions of the Stokes parameter for linear polarization ($Q$) to be very sensitive to the geometric distribution of the dust (Figure~\ref{fig_kernel_q00}). As a result, dust distributions that predict similar light curves can be more efficiently distinguished if detailed time evolution of polarization can be acquired  (Figure~\ref{fig_lcs_polar}). Our study shows that a time sequence of broadband polarimetry is a more powerful probe for determining the dust geometry than detailed spectropolarimetry but with less time coverage. We also compared the results between our studies and those of \cite{Bulla2018MNRAS.479.3663B}, and found that shell models with considerably different distance scales can fit the time dependence of the $E(B-V)$ curves (Figure~\ref{fig_color_06x}); we argue that the location of the dust grains responsible for any time-varying reddening of SNe~Ia cannot be determined reliably based on photometric optical data alone. Late-time polarimetry, especially broadband polarimetry from a few months to over a year, can be of great value in setting limits on the elusive CS dust around SNe~Ia.

\acknowledgments

This work is supported by the National Natural Science Foundation of China (11761141001) and Key Research Program of Frontier Sciences of Chinese Academy of Sciences (QYZDY-SSW-SLH010). X. Wang is supported by National Natural Science Foundation of
China (NSFC grants 12033003 and 11633002). This work is partially supported by the Scholar Program of Beijing Academy of Science and Technology (DZ: BS202002). We thank M. Bulla, Lingzhi Wang, and Y. Yang for sharing the data used in this paper.

\bibliography{bibtex}{}
\bibliographystyle{aasjournal}



\end{document}